\documentclass{gGAF2e}
\usepackage{multiequations}

\newcommand*{\bme}{\begin{multiequations}}
\newcommand*{\eme}{\end{multiequations}}
\newcommand*{\be}{\begin{equation}}
\newcommand*{\ee}{\end{equation}}

\newcommand*{\te}{\tripleequation}

\begin{document}
\jvol{00} \jnum{00} \jyear{2012} 

\markboth{\rm A.R. YEATES}{\rm GEOPHYSICAL \&  ASTROPHYSICAL FLUID DYNAMICS}


\title{On the limitations of magneto-frictional relaxation}

\author{A.~R. YEATES$^{\ast}$\thanks{$^\ast$Corresponding author. Email: anthony.yeates@durham.ac.uk
\vspace{6pt}}\\\vspace{6pt}
Department of Mathematical Sciences, Durham University, Durham, DH1 3LE, UK\\
\vspace{6pt}\received{\today} }

\maketitle

\begin{abstract}
The magneto-frictional method is used in solar physics to compute both static and quasi-static models of the Sun's coronal magnetic field. Here, we examine how accurately magneto-friction (without fluid pressure) is able to predict the relaxed state in a one-dimensional test case containing two magnetic null points. Firstly, we show that relaxation under the full ideal magnetohydrodynamic equations in the presence of nulls leads necessarily to a non-force-free state, which could not be reached exactly by magneto-friction. Secondly, the magneto-frictional solutions are shown to lead to breakdown of magnetic flux conservation, whether or not the friction coefficient is scaled with magnetic field strength. When this coefficient is constant, flux is initially conserved, but only until discontinuous current sheets form at the null points. In the ensuing weak solution, we show that magnetic flux is dissipated at these current sheets. The breakdown of flux conservation does not occur for an alternative viscous relaxation scheme.

\end{abstract}

\section{Introduction} \label{sec:intro}

Magneto-friction (hereafter abbreviated to MF) is a computational method for obtaining force-free magnetic equilibria, whereby one retains the (ideal) magnetohydrodynamic (MHD) induction equation
\begin{equation}
    \frac{\upartial \bm{B}}{\upartial t} = {\bm\nabla}\times\big(\bm{u} \times \bm{B}\big),
    \label{eqn:induc}
\end{equation}
but specifies the frictional velocity
\begin{equation}
    \bm{u} = \nu^{-1}\bm{J}\times\bm{B},
    \label{eqn:umf}
\end{equation}
rather than solving the full fluid equations. Here $\bm{B}$ is the magnetic field and $\bm{J}=\mu_0^{-1}{\bm\nabla}\times\bm{B}$ is the current density. This form of velocity changes equation \eqref{eqn:induc} from hyperbolic to (degenerate) parabolic type \citep{craig1986}, and indeed the magnetic energy in a volume $V$ satisfies
\begin{equation}
    \frac{\mathrm{d}}{\mathrm{d}t}\int_V\frac{B^2}{2\mu_0}\,\mathrm{d}V = -\int_V\nu^{-1}|\bm{J}\times\bm{B}|^2\,\mathrm{d}V - \oint_{\upartial V}\nu^{-1}B^2\bm{J}\times\bm{B}{\bm\cdot}\,\mathrm{d}\bm{S}.
    \label{eqn:W}
\end{equation}
Thus energy is dissipated monotonically within the volume until a force-free state $\bm{J}\times\bm{B}$ is reached.

The basic method dates back at least to \citet{chodura1981}, and was adopted (with $\nu=1$) by \citet{craig1986}, who called it a ``fictitious fluid'' and introduced a Lagrangian numerical method with the aim of better preserving magnetic flux conservation. \citet{yang1986} used a friction coefficient of the form $\nu\sim B^2$, which has the effect of accelerating the relaxation in regions of weak $B$. They observed a change in magnetic topology during the relaxation, but attributed this to the effect of numerical diffusion.

Subsequent authors have used MF in different ways to model the magnetic field in the solar corona, taking it to be force-free to first approximation \citep{wiegelmann2021}. For example, \citet{klimchuk1992} wrote $\bm{B}$ in terms of Euler potentials and used these to specify the field line connectivity on the lower boundary (the solar photosphere). \citet{roumeliotis1996} developed a ``stress-and-relax'' method for computing force-free equilibria, where the magnetic field on the lower boundary is updated step-by-step toward an observed vector magnetogram, with the coronal field responding step-by-step by MF. This technique has been successfully used for modelling solar active regions \citep{valori2005, valori2010, guo2016}. 

MF has also been used (with sufficiently small $\nu$) to model the quasi-static response of the coronal magnetic field to slowly evolving boundary conditions, rather than computing a single equilibrium. This idea was introduced by \citet{vanballegooijen2000}, who modelled the evolution of the large-scale coronal magnetic field in response to both large-scale solar surface motions (primarily differential rotation) and small-scale (supergranular) convection, which they parametrised by a surface diffusion term. They showed that this simple model can capture the formation of sheared filament channels, and even the sudden eruption of twisted magnetic structures due to loss of equilibrium \citep[e.g.,][]{mackay2012, lowder2017, bhowmik2021}. The MF model has also been applied with resolved small-scale boundary driving, using either an imposed convective velocity field \citep{meyer2016} or an imposed electric field constrained by observed magnetograms \citep{mackay2011, cheung2012, pomoell2019, hoeksema2020, yardley2021}. The electric-field driven MF simulations have been successful at reproducing the formation of observed non-potential magnetic structures within individual active regions, and have also performed favourably in reconstructing the injected magnetic energy and relative helicity when compared directly to a full-MHD flux emergence simulation \citep{toriumi2020}.

In this paper, our aim is to explore how well MF is able to reproduce the equilibrium state that would be predicted by relaxation under the full ideal MHD equations. \citet{goldstraw2018} have recently published just such a test for a setup that models a solar coronal loop whose magnetic field lines are sheared by footpoint displacements at either end. They find that MF gives an excellent approximation of the quasi-static sequence of MHD equilibria for small plasma-beta, defined as $\beta = 2\mu_0 p/B^2$, where $p$ is the fluid pressure. For large $\beta$, their MF result departs from the MHD solution because it neglects the effects of pressure. Here, we build on that study by considering a magnetic field containing null points ($B=0$). Near these points, which are common in the coronal models mentioned above, the $\beta$ parameter is necessarily large and this raises the question of whether MF -- as given by equation \eqref{eqn:umf} -- will be successful at reproducing the ideal MHD relaxation of such magnetic fields.

It is important to note that some implementations of the MF method do include an additional fluid pressure term in equation \eqref{eqn:umf}, writing $\bm{u}=\nu^{-1}(\bm{J}\times\bm{B} - {\bm\nabla} p)$. These include the original calculations of \citet{chodura1981}, those of \citet{linardatos1993}, and the Lagrangian schemes of \citet{craig2005} or \citet{candelaresi2015}. There are also implementations that add an MF term to a momentum equation where the inertial terms are retained \citep{hesse1993,candelaresi2015}. As such, our conclusions will apply only to the MF model without  additional pressure or inertial terms, which is nevertheless widely used in solar physics.

Some particular criticisms of the MF method were raised by \citet{low2013}. The first important objection is that null points cannot move during magneto-friction, since at such points equation \eqref{eqn:umf} implies that $\bm{u}=\bm{0}$. This suggests that the method will fail if the null points need to move during the relaxation, as will be the case for our test in this paper. The second objection is that discontinuous current sheets (jumps in $\bm{B}$ with infinite $\bm{J}$) will form in finite time, so that the method breaks down. We shall return to both of these important points below. Again, it should be noted that both of these criticisms apply to the MF model without plasma pressure, but need no longer apply if a pressure term is added to the velocity.

The outline of the paper is as follows. Section \ref{sec:setup} describes our test setup, along with the ``ground truth'' ideal MHD solution. In order to be sure that our results are not affected by numerical diffusion, we adopt a simple one-dimensional setup, inspired by \citet{bajer2013}. In section \ref{sec:mf}, we present one-dimensional MF results for different forms of $\nu$, and we conclude in section \ref{sec:dis}  with a discussion of the implications for future modelling.

\section{Test setup} \label{sec:setup}

\begin{figure}
    \begin{center}
    \includegraphics{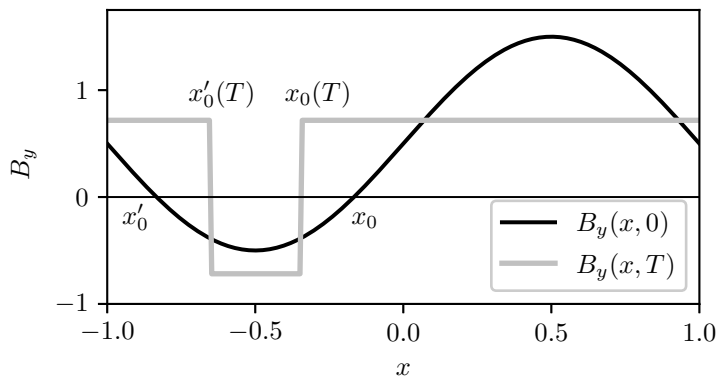}
    \caption{The initial magnetic field for our test setup (dark line). The feint line shows the state of minimum magnetic energy, discussed in section \ref{sec:min}. The positions of the null points in the former and current sheets in the latter are indicated.}
    \label{fig:init}
    \end{center}
\end{figure}

We will consider the relaxation of a magnetic field that initially has the (non-equilibrium) form
\bme
\label{eqn:init}
\begin{equation}
    B_x(x,0) = B_z(x,0)=0, \hskip 10mm B_y(x,0) = \tfrac12 + \sin(\pi x),
\end{equation}
\eme
on the domain $-1\leq x \leq 1$. For simplicity we will assume periodic boundary conditions on all variables. This initial magnetic field is shown by the solid curve in figure~\ref{fig:init}. The offset of $1/2$ is added so that the magnetic null points, which are initially at $x=x_0=-1/6$ and $x=x_0'=-5/6$, will move during the relaxation, allowing a fairer test of the MF method.

\subsection{Ground truth: full MHD} \label{sec:mhd}

\begin{figure}
    \begin{center}
    \includegraphics[width=\textwidth]{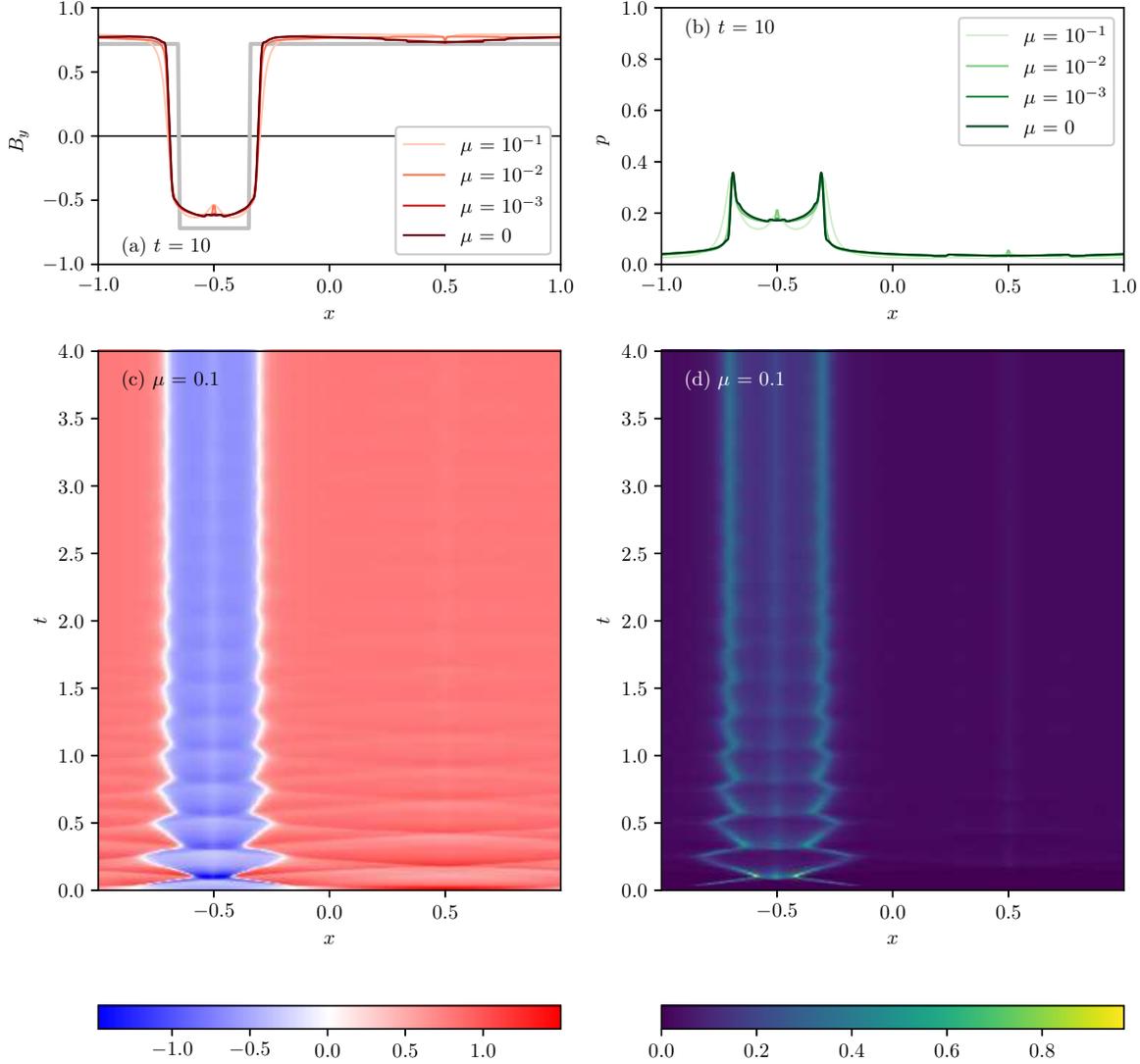}
    \caption{Illustration of the full MHD relaxation. Panels (a) and (b) show $B_y$ and $p$ at $t=10$ for several values of $\mu$, while panels (c) and (d) show the time evolution up to $t=4$ of $B_y$ and $p$ for the case $\mu=10^{-1}$. The thick gray line in (a) is the minimum energy solution in section \ref{sec:min}. (Colour online)}
    \label{fig:stackplot}
    \end{center}
\end{figure}

Here we illustrate the solution of the full ideal MHD equations with our initial condition~\eqref{eqn:init}, where we take the initial pressure and density to be constant, with $p(x,0)=0$ and $\rho(x,0)=10^{-2}$. Thus we begin with a low-beta plasma away from the locations where $B_y\approx 0$, consistent with the solar coronal plasmas where magneto-friction is typically applied. We assume an adiabatic ideal gas with polytropic index $\gamma=5/3$, so that the ideal MHD equations for this initial condition reduce to
\begin{subequations}
\label{eqn:mhd}
\begin{align}
\frac{\upartial{\rho}}{\upartial{t}} &= -\frac{\upartial{}}{\upartial{x}}\big(\rho u_x\big),\label{eqn:rho}\\
\rho\left(\frac{\upartial{u_x}}{\upartial{t}} + u_x\frac{\upartial{u_x}}{\upartial{x}}\right) &= -\frac{\upartial}{\upartial{x}}\left( p + \frac{B_y^2}{2\mu_0}\right) + \mu\frac{\upartial{^2u_x}}{\upartial{x^2}},\label{eqn:u}\\
\frac{\upartial{p}}{\upartial{t}} + u_x\frac{\upartial{p}}{\upartial{x}}  &= -\gamma p\frac{\upartial{u_x}}{\upartial{x}} + (\gamma-1)\mu\left(\frac{\upartial{u_x}}{\upartial{x}}\right)^{\!2},\label{eqn:p}\\
\frac{\upartial{B_y}}{\upartial{t}} &= -\frac{\upartial{}}{\upartial{x}}\big(u_xB_y\big). \label{eqn:ind}
\end{align}
\end{subequations}
A uniform viscosity $\mu$ is included, together with corresponding viscous heating term in the pressure equation \eqref{eqn:p} to ensure conservation of total energy $E = E_{\rm kin} + E_{\rm int} + E_{\rm mag}$, where the kinetic, internal, and magnetic energies are
\bme
\te
\begin{equation}
E_{\rm kin} = \int_{-1}^1\frac{\rho u_x^2}{2}\,\mathrm{d}x, \hskip10mm 
    E_{\rm int} = \int_{-1}^1\frac{p}{\gamma-1}\,\mathrm{d}x,  \hskip10mm 
    E_{\rm mag} = \int_{-1}^1\frac{B_y^2}{2\mu_0}\,\mathrm{d}x.\hskip10mm 
\end{equation} 
\eme
We solve equations \eqref{eqn:mhd} numerically with the ATHENA code \citep{stone2008}, using the HLLD Riemann solver with third-order reconstruction on a uniform grid of 1024 points in the $x$-direction. We set $\mu_0=1$ so that the maximum Alfv\'en speed is order $\sqrt{\rho^{-1}}=10$ in our units. Results are illustrated in figures~\ref{fig:stackplot}(a,b) for the inviscid case $\mu=0$ as well as three increasing values of viscosity $\mu$. Figures~\ref{fig:stackplot}(c,d) show the full time evolution for the single case $\mu=10^{-1}$.

First, consider the evolution of $\bm{B}$. When the evolution begins, Alfv\'en waves are immediately launched in opposite directions from the region of peak $B_y(x,0)$ around $x=1/2$. These are seen as diagonal fronts in figure~\ref{fig:stackplot}(c), moving at speed approximately 10. These waves interact each time they cross the domain, as the magnetic field relaxes to a lower energy state.

\begin{figure}
    \begin{center}
    \includegraphics[width=0.8\textwidth]{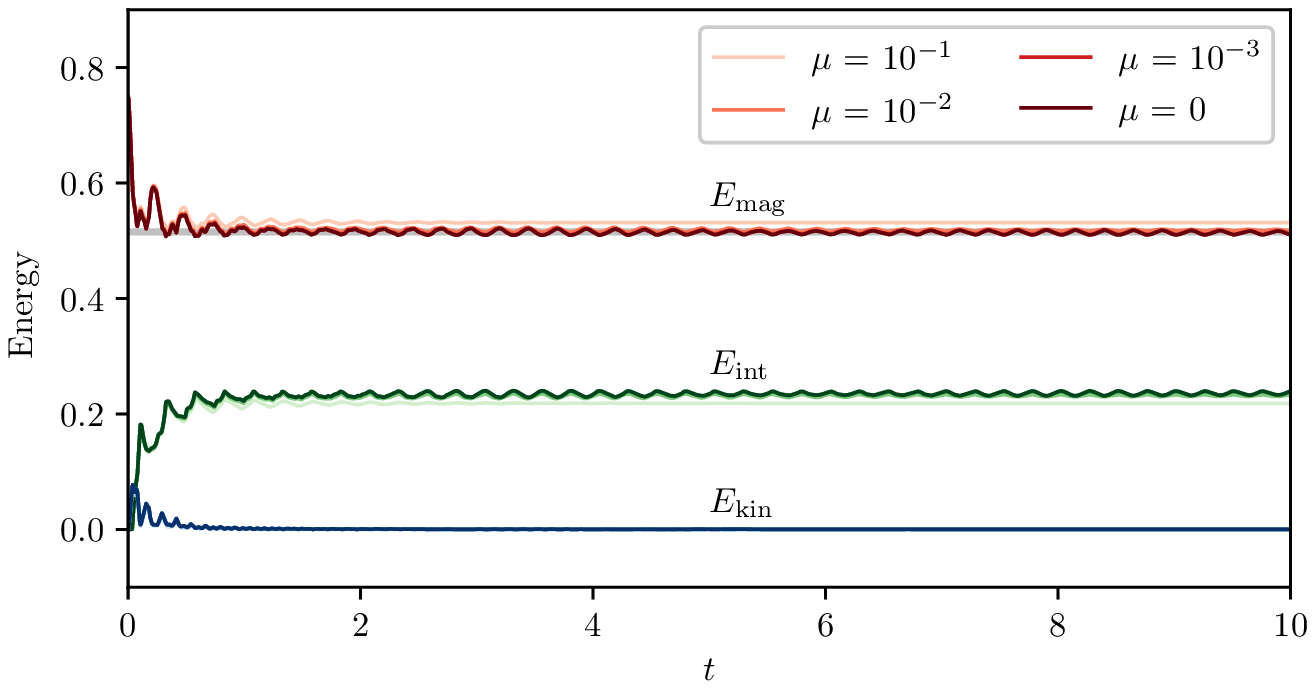}
    \caption{Time evolution of the three components of the total energy in the full MHD relaxation, for different values of viscosity $\mu$. (For each component, lighter coloured curves indicate larger $\mu$.) The thick gray horizontal line (just visible) indicates the minimum magnetic energy under an ideal relaxation with no further constraints (section \ref{sec:min}). (Colour online)}
    \label{fig:energies}
    \end{center}
\end{figure}

Figure \ref{fig:energies} shows that the magnetic energy reduces from its initial value of $0.75$ to approximately $0.52$ during this initial relaxation phase, which lasts until around $t=1$. After this time, the kinetic energy is very small, even for $\mu=0$ although in that case some residual waves remain, as evidenced by the oscillations at the Alfv\'en frequency in $E_{\rm mag}$. These oscillations are damped by viscosity, and for $\mu=10^{-1}$ this happens before $t=4$, as seen in figure~\ref{fig:stackplot}(c).

It is clear from figure~\ref{fig:energies} that the magnetic energy lost during the relaxation goes entirely into internal energy, as must be the case if a stationary equilibrium is reached with $E_{\rm kin}=0$, owing to conservation of total energy. This internal energy is manifested in the relaxed state by a non-zero fluid pressure, as shown in figure \ref{fig:stackplot}(b). The pressure is strongest in the regions where $B_y$ changes rapidly, consistent with this being a magnetostatic equilibrium, which in our one-dimensional system amounts to a total pressure balance
\begin{equation}
    \frac{\upartial{}}{\upartial{x}}\left(p + \frac{B_y^2}{2\mu_0}\right) = 0.
\end{equation}
Thus in the final equilibrium the plasma-beta is quite variable: it is small in the regions where $B_y$ was initially strong and so decreased, while it is larger (0.2 to 0.4) in the region where the fluid has been compressed in order to increase $B_y$. This conclusion is little changed by the presence of viscosity (at least for the $\mu$ values considered here).

\subsection{State of minimum magnetic energy} \label{sec:min}

Since the MF method will be ignoring the fluid equations (\ref{eqn:rho}-c),
it is useful to compare the full MHD solution to the state of minimum $E_{\rm mag}$ constrained only by the induction equation \eqref{eqn:ind}. This would be a state of uniform magnetic pressure, $B_y^2 = \textrm{constant}$, and for our initial condition \eqref{eqn:init}, conservation of total unsigned flux shows that
\begin{equation}
    |B_y| = \frac12\int_{-1}^1|B_y(x,0)|\,\mathrm{d}x = \frac16 + \frac{\sqrt{3}}{\pi}.
\end{equation}
Since the magnetic topology is preserved, the two null points must survive and become discontinuous current sheets in the relaxed state, at which $B_y$ changes sign \citep[cf.][]{bajer2013}. From conservation of flux between the nulls, and symmetry, it follows that these current sheets would be located at
\bme
\te
\begin{equation}
  x_0(T) = \frac{-1 +\varDelta}{2}, \hskip 10mm  x_0'(T) = \frac{-1 - \varDelta}{2}, \hskip 10mm \varDelta = \frac{6\sqrt{3}-2\pi}{6\sqrt{3} + \pi}.\hskip 5mm
\end{equation}
\eme
This minimum $E_{\rm mag}$ magnetic field is shown by the feint line in figure~\ref{fig:init}, repeated in figure~\ref{fig:stackplot}(a) for comparison to the full MHD solution. It is evident that the finite pressure in the full MHD solution modifies the shape of the equilibrium and smooths out the current sheets to some extent. Nevertheless, the magnetic energy in the MHD solution reaches close to that of the minimum $E_{\rm mag}$ solution, which is
\begin{equation}
    \overline{E}_{\rm mag} = \frac{1}{36} + \frac{\sqrt{3}}{3\pi} + \frac{3}{\pi^2} \approx 0.516.
\end{equation}
This is shown by the feint grey horizontal line in figure~\ref{fig:energies}.

\section{Magneto-frictional results} \label{sec:mf}

\begin{figure}
    \centering
    \includegraphics[width=\textwidth]{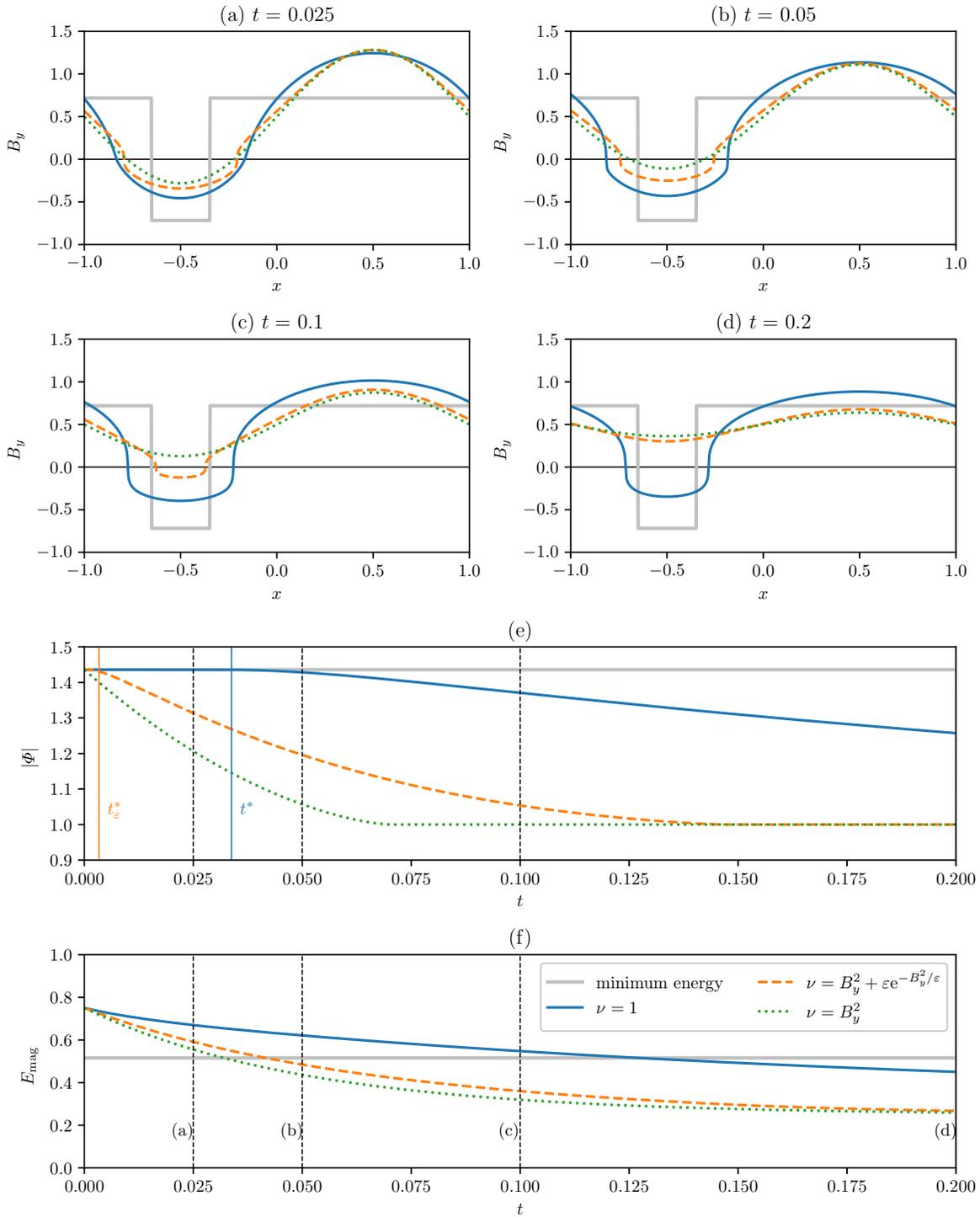}
    \caption{MF results for the test case with each form of $\nu$. Panels (a) to (d) show the solutions at selected times, according to the legend in panel (f). The thick gray line shows the minimum energy solution from section \ref{sec:min}. Panel (e) shows the total unsigned magnetic flux as a function of time, and panel (f) shows the magnetic energy. Vertical dashed lines in (e) and (f) indicate the times of panels (a)-(d), and vertical solid lines in (e) indicate the breakdown times $t^*$ and $t^*_\varepsilon$ as described in the text. We take $\varepsilon=10^{-1}$. (Colour online)}
    \label{fig:mf}
\end{figure}

For a one-dimensional magnetic field of the form $B_x=B_z=0$, $B_y=B_y(x,t)$, the frictional velocity \eqref{eqn:umf} reduces to
\bme
\label{eqn:u1d}
\begin{equation}
    u_x = -\frac{1}{\nu}\frac{\upartial{}}{\upartial{x}}\left(\frac{B_y^2}{2\mu_0}\right), \hskip 10mm u_y = u_z = 0,
\end{equation}
\eme
so equation \eqref{eqn:induc} takes the form
\begin{equation}
        \frac{\upartial{B_y}}{\upartial{t}} = \frac{1}{\mu_0}\frac{\upartial{}}{\upartial{x}}\left(\frac{B_y^2}{\nu}\frac{\upartial{B_y}}{\upartial{x}}\right). \label{eqn:mf1d}
\end{equation}
The behaviour of this equation depends on the chosen form of the friction coefficient $\nu$. Overall scaling of $\nu$ is simply equivalent to scaling the time variable, so we only need to consider its functional form. In the following subsections, we will consider three illustrative cases:
\begin{enumerate}
    \item $\nu = B_y^2$;
    \item $\nu = 1$;
    \item $\nu = B_y^2 + \varepsilon\mathrm{e}^{-B_y^2/\varepsilon}$, where $\varepsilon$ is a small constant ($10^{-1}$ in our computations).
\end{enumerate}
These cover the basic forms that have been used by different authors in the literature. The results of these three test cases are shown in figure~\ref{fig:mf}, and discussed in the following subsections. For cases (ii) and (iii), equation \eqref{eqn:mf1d} was solved with the Crank-Nicolson method, treating the nonlinearity by Picard iteration (see Appendix \ref{sec:app} for details). Results shown in figure~\ref{fig:mf} used mesh resolution $n_x=1024$. The unsigned magnetic flux in figure~\ref{fig:mf}(e) is defined as
\begin{equation}
    |\varPhi| = \int_{-1}^1|B_y|\,\mathrm{d}x.
    \label{eqn:unflux}
\end{equation}

\subsection{Case (i) -- linear diffusion}

As noted in section \ref{sec:intro}, applications of MF have commonly set $\nu=B^2$, in order to speed up relaxation in regions of low $B$. In our one-dimensional case, this reduces \eqref{eqn:mf1d} to the linear diffusion equation
\begin{equation}
    \frac{\upartial{B_y}}{\upartial{t}} = \frac{1}{\mu_0}\frac{\upartial{}^2B_y}{\upartial{x^2}}, \label{eqn:diff}
\end{equation}
whose solution remains smooth for all time. For the test case \eqref{eqn:init}, the solution is precisely $B_y(x,t) = 1/2 + \sin(\pi x)\exp(-\pi^2t)$, shown by the green dotted lines in figure~\ref{fig:mf}.

Unfortunately, equation \eqref{eqn:diff} no longer conserves (unsigned) magnetic flux, as  evidenced by the green dotted line in figure~\ref{fig:mf}(e). (After approximately $t=0.07$ unsigned flux is conserved because the two null points have been completely eliminated.) Thus it violates a fundamental property of ideal relaxation. Although we are still solving the induction equation \eqref{eqn:ind}, there is no contradiction because the frictional velocity from \eqref{eqn:u1d} is
\begin{equation}
    u_x = -\frac{1}{\mu_0B_y}\frac{\upartial{B_y}}{\upartial{x}},
\end{equation}
which is singular at null points so that flux conservation breaks down \citep[e.g.][]{wilmotsmith2005}. Not surprisingly, the magnetic energy of this solution in figure~\ref{fig:mf}(f) reaches far below that of the ideal minimum energy state (in the limit, it tends to $1/4$, approximately half that of the ideal minimum).

More generally, it would be only the perpendicular current that would be diffused with $\nu = B^2$. To see this, note that, in three dimensions, setting $\nu=B^2$ in equations \eqref{eqn:induc} and \eqref{eqn:umf} would give
\bme
\begin{equation}
    \frac{\upartial{\bm{B}}}{\upartial{t}} = -{\bm\nabla}\times\bm{J}_\perp, \hskip 10mm \bm{J}_\perp = \bm{J} - \frac{\bm{J}{\bm\cdot}\bm{B}}{B^2}\bm{B}.
\end{equation}
\eme
Nevertheless, our one-dimensional test suggests that the method is unable in general to accurately reconstruct an ideal relaxed state if null points are present. For example, we propose that this diffusion, rather than numerical error, could have been responsible for the change in magnetic topology observed in two dimensions by \citet{yang1986}.

\subsection{Case (ii) -- ambipolar diffusion}

Next, consider the case of constant $\nu$. In that case, equation \eqref{eqn:mf1d} is equivalent to the induction equation for a partially-ionised plasma under ambipolar diffusion and vanishing plasma velocity \citep{brandenburg1994, hoyos2010}. It is a particular case of the porous medium equation \citep{vazquez2006}. An important property of this equation is that discontinuous current sheets develop at finite time from null points $B_y=0$ in the initial state. We can predict the time of this breakdown by considering the slope $w_{x_0}(t)= (\upartial B_y/\upartial x)_{x_0}$ at a null point $x=x_0$ \citep{low2013}. As long as the solution remains smooth, the null point cannot move because $u_x(x_0,t)=0$ by equation \eqref{eqn:u1d}. So differentiating \eqref{eqn:mf1d} with respect to $x$ and setting $B_y=0$ shows that
\begin{equation}
    \frac{\mathrm{d}w_{x_0}}{\mathrm{d}t} = \frac{2}{\mu_0}w_{x_0}^3
\end{equation}
and hence the slope at the null, $w_{x_0}(t)$, obeys
\begin{equation}
    w_{x_0}^2(t) = \frac{w_{x_0}^2(0)}{1 - 4w_{x_0}^2(0)t}.
\end{equation}
It follows that the smooth solution will break down and a current sheet will form at time $t^* = [4w_{x_0}^2(0)]^{-1}$. Similar behaviour is expected in higher dimensions, and indeed finite-time formation of a discontinuous current sheet at a three-dimensional null point is strongly suggested by the numerical results of \citet{pontin2012}, who also use MF with constant $\nu$. Such breakdown of smoothness is not in itself a problem in our test case, where we know that the minimum energy state under ideal relaxation neglecting fluid pressure would have discontinuous current sheets. Incidentally, \citet{craig2005} show that adding a pressure gradient term to the magneto-frictional velocity \eqref{eqn:umf} can prevent the formation of discontinous current sheets in one-dimensional configurations (see their Appendix A). But in higher dimensions, \citet{pontin2005} demonstrate that adding a pressure term is insufficient to prevent current-sheet collapse, since the Lorentz force is not generally irrotational.

After $t^*$, which takes the value $t^*=(3\pi^2)^{-1}$ for our test case, we can continue to follow the evolution as a weak solution, since \eqref{eqn:mf1d} has the form of a conservation law. However, two properties that held while the solution was smooth are no longer true: the null points can now move, and the unsigned magnetic flux is no longer conserved. We now discuss each of these in turn.

\begin{figure}
    \centering
    \includegraphics[width=\textwidth]{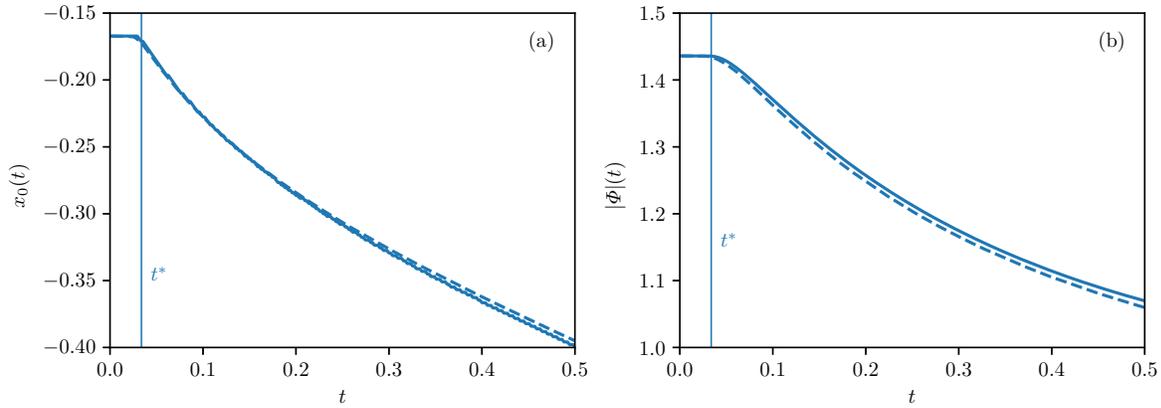}
    \caption{Verification that the numerical MF solution with $\nu=1$ satisfies the formulae \eqref{eqn:rankine} and \eqref{eqn:diss}. The solid curves show (a) the position of the right-hand null/current sheet and (b) the total unsigned magnetic flux. The dashed curves show predictions of these quantities found by estimating the jumps in \eqref{eqn:rankine} and \eqref{eqn:diss} numerically and integrating these expressions in time. (Colour online)}
    \label{fig:jumps}
\end{figure}

It is evident from the time snapshots in figure~\ref{fig:mf}(a--d) that, once the null points have degenerated into discontinuous current sheets, these start to move. Their speed of movement is given by the standard Rankine-Hugoniot condition,
\begin{equation}
    \frac{\mathrm{d}x_0}{\mathrm{d}t} = -[B_y]^{-1}_{x_0(t)}\left[\frac{B_y^2}{\nu}\frac{\upartial{B_y}}{\upartial{t}} \right]_{x_0(t)},
    \label{eqn:rankine}
\end{equation}
where square brackets denote the limiting jump in a quantity across the current sheet $x=x_0(t)$. Note that the right-hand side would vanish if the ideal minimum energy state were reached, although it is not actually reached. Figure~\ref{fig:jumps}(a) shows that our numerical solution recovers the correct speed for the current sheet that starts at $x_0(t^*)=-1/6$. It is clear that the objection of \citet{low2013} to MF -- that null points cannot move -- holds only while the solution remains smooth, and is not relevant in the long term.

Next we consider the breakdown in conservation of unsigned flux after $t^*$. This is clearly evident in figure~\ref{fig:mf}(e), and is not caused by numerical error. The fact that ambipolar diffusion leads to loss of magnetic flux after the formation of current sheets was shown by \citet{hoyos2010} for a $B_y$ profile with a single step. They called it ``reconnection in the absence of ohmic diffusivity''. For our periodic domain with discontinuous current sheets, the dissipation rate of unsigned flux may be expressed in terms of jumps at the current sheets. In our example with two current sheets $x_0(t)$ and $x_0'(t)$, careful evaluation of the integral \eqref{eqn:unflux} shows -- after some algebra -- that 
\begin{subequations}
\begin{align}
\frac{\mathrm{d}|\varPhi|}{\mathrm{d}t} &= [B_y]^{-1}_{x_0'(t)}\left[\frac{2B_y^3}{\nu}\frac{\upartial{B_y}}{\upartial{x}}\right]_{x_0'(t)} - [B_y]^{-1}_{x_0(t)}\left[\frac{2B_y^3}{\nu}\frac{\upartial{B_y}}{\upartial{x}}\right]_{x_0(t)}\\
&= -4[B_y]^{-1}_{x_0(t)}\left[\frac{B_y^3}{\nu}\frac{\upartial{B_y}}{\upartial{x}}\right]_{x_0(t)}, \label{eqn:diss}
\end{align}
\end{subequations}
where in the second line we used the symmetry of the setup. Although we cannot evaluate these quantities analytically, equation \eqref{eqn:diss} gives us a relation that we can use to check that the dissipation observed in the numerical solution is accurate. Indeed, figure~\ref{fig:jumps}(b) shows that our numerical method obeys this relation to reasonable accuracy, despite the difficulty in estimating the limiting quantities on either side of the current sheet from the numerical solution. This arises since $B_y$ is not uniform on each side of the current sheet but varies like $(x-x_0)^{1/3}$ \citep{brandenburg1994}. In conclusion, it is clear that the dissipation of unsigned flux after $t^*$ is a real phenomenon. This clearly prevents the method from recovering the correct minimum-energy state, although it is evident from figure~\ref{fig:mf}(e) that the unwanted dissipation is much slower than for linear diffusion.

\subsection{Case (iii) -- linear diffusion with limiting}

\begin{figure}
    \centering
    \includegraphics[width=0.6\textwidth]{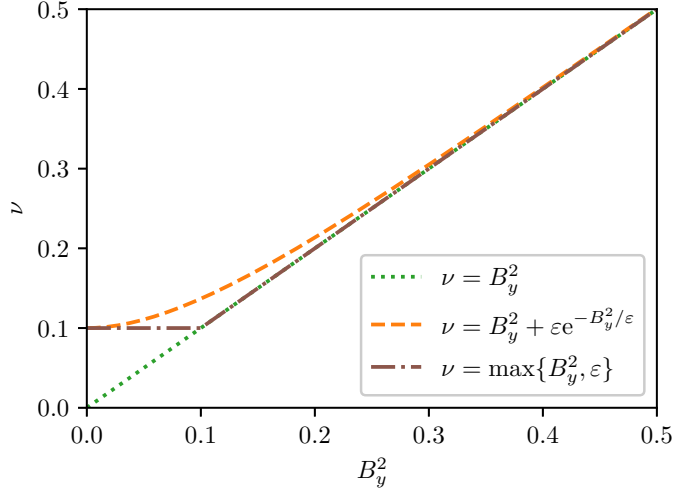}
    \caption{Functional forms of the friction coefficient with limiting, for $\varepsilon=10^{-1}$, compared to the linear diffusion case. (Colour online)}
    \label{fig:nulim}
\end{figure}

The final friction coefficient we consider is $\nu = B_y^2 + \varepsilon\mathrm{e}^{-B_y^2/\varepsilon}$. This functional form is plotted in figure~\ref{fig:nulim}, and is seen to limit the minimum value of $\nu$ to $\varepsilon$, while keeping $\nu\approx B_y^2$ away from null points, as with linear diffusion. Such a limit is typically (albeit tacitly) employed in numerical calculations; for example, \citet{yeates2014} use the simpler non-smooth form $\nu = \max\{B_y^2,\varepsilon\}$, also shown in figure~\ref{fig:nulim}. The parameter $\varepsilon$ would be set to a small number. The resulting equation is
\begin{equation}
            \frac{\upartial{B_y}}{\upartial{t}} = \frac{1}{\mu_0}\frac{\upartial{}}{\upartial{x}}\left(\frac{B_y^2}{B_y^2 + \varepsilon\mathrm{e}^{-B_y^2/\varepsilon}}\frac{\upartial{B_y}}{\upartial{x}}\right). \label{eqn:lim}
\end{equation}
This limited form of $\nu$ is interesting mathematically because it prevents the frictional velocity \eqref{eqn:u1d} from being singular at null points, thus preserving flux conservation. However, there is a catch: near null points, the equation looks like the ambipolar diffusion case (constant $\nu$). Thus we might expect the finite-time breakdown of smoothness and formation of discontinuous current sheets at nulls. Indeed we can apply the argument of \citet{low2013} to equation \eqref{eqn:lim} to see -- after some algebra -- that a smooth solution with a null point will break down at $t^*_\varepsilon = \varepsilon[4w_{x_0}^2(0)]^{-1}$, where $w_{x_0}(0)$ is again the initial slope of $B_y$ at the null. In particular, for $\varepsilon<1$, this will occur more rapidly than for the ambipolar diffusion case.

The orange dashed curves in figure~\ref{fig:mf} show the numerical solution with the limiting form of $\nu$ for $\varepsilon = 10^{-1}$, and indeed a discontinuous current sheet forms already at $t=t^*_\varepsilon$. As for the constant $\nu$ solution, the null points remain stationary and flux is conserved until this breakdown, but both properties are violated thereafter. It should be noted that a large value of $\varepsilon=10^{-1}$ was used here for illustration. As $\varepsilon$ is reduced, $t^*_\varepsilon$ gets earlier and earlier, while the current sheets at any fixed later time have less and less of a jump in $B_y$, with the solution tending to the linear diffusion case as $\varepsilon\to 0$. For the small $\varepsilon$ typically used in simulations to avoid division by zero, the solution would be essentially indistinguishable from the linear diffusion case, despite the fact that the frictional velocity is initially non-singular.

\section{Discussion}  \label{sec:dis}

We have shown that, when a magnetic null point is present, magneto-frictional (MF) relaxation is unable to accurately predict the relaxed state that would be obtained by an ideal MHD evolution. We have demonstrated two separate reasons for this. Firstly, as shown in section~\ref{sec:mhd}, a relaxation under the full MHD equations starting from a uniform pressure will -- at least in our one-dimensional example -- build up a non-trivial distribution of fluid pressure, owing to the fact that magnetic energy is transformed to internal energy. In particular, fluid pressure accumulates at the magnetic null points in order to reach total pressure balance. Thus the true equilibrium will be magnetostatic rather than force-free. In effect, the plasma-beta is high near the null points, even if it is low elsewhere. This motivates the use of relaxation schemes that include fluid pressure, while still avoiding the need for small timesteps to resolve MHD waves. As discussed, one approach is to modify the MF velocity to $\bm{u}=\nu^{-1}(\bm{J}\times\bm{B} - {\bm\nabla} p)$. To determine $p$ in two-dimensional simulations, \citet{linardatos1993} used 
the requirement that ${\bm\nabla}{\bm\cdot}\bm{u}=0$ \citep[see also][]{moffattbook}. Unfortunately, this would not work for our one-dimensional test case, where the relaxation velocity is necessarily compressible. Instead, one could write ${\bm\nabla} p = \beta{\bm\nabla}\rho$, and evolve $\rho$ by mass conservation \citep{craig2005,candelaresi2015}.

A more significant limitation of MF without pressure is that it fails to respect conservation of unsigned magnetic flux, which would follow from the ideal induction equation \eqref{eqn:ind} provided that $\bm{u}$ remains smooth. The problem is that $\bm{u}$ does not remain smooth for any of the forms of MF in common use. When $\nu=B_y^2$, our one-dimensional problem reduces to a linear diffusion equation. When $\nu$ is constant, or becomes constant near to a null point, then $\bm{u}$ is initially smooth but this breaks down at a finite time. In itself, we would argue \citep[in contrast to][]{low2013} that this formation of discontinuous current sheets is not a problem, since the state of minimum magnetic energy when fluid pressure is neglected is a discontinuous one (section \ref{sec:min}). However, the ensuing weak solution obtained does not conserve unsigned flux, so does not evolve toward this expected solution.
For one-dimensional configurations such as that investigated here, \citet{craig2005} show that adding plasma pressure can also prevent the formation of a discontinuous current sheet, and hence also the breakdown of flux conservation.

\begin{figure}
    \centering
    \includegraphics[width=\textwidth]{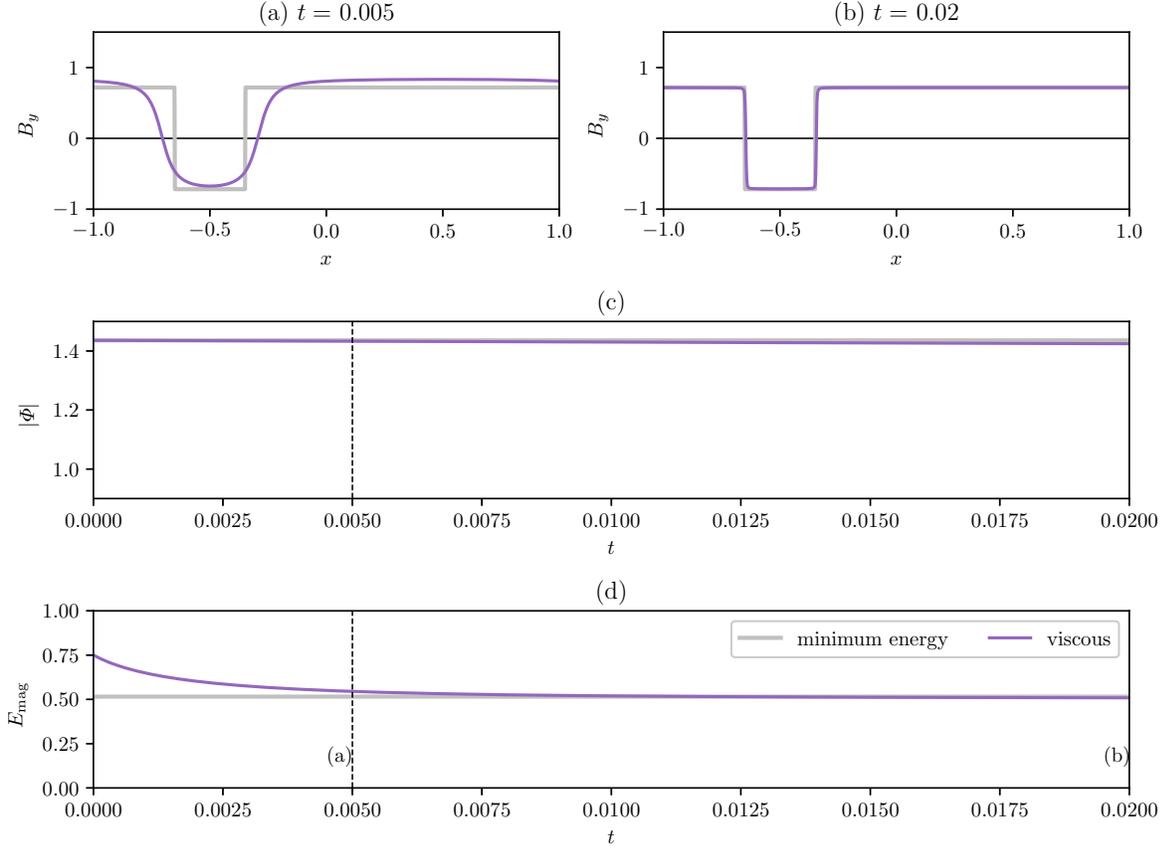}
    \caption{Result of viscous relaxation (with $\mu=1$) for the test case. As in figure~\ref{fig:mf}, panels (a) and (b) show snapshots of the solution during relaxation, while (c) and (d) show the unsigned magnetic flux and magnetic energy, respectively. The thick gray line shows the minimum energy solution from section \ref{sec:min}. (Colour online)}
    \label{fig:visc}
\end{figure}

To end this paper on a more positive note, we point out an alternative method that is able to avoid the second limitation (flux non-conservation) and reach the expected minimum energy state from section \ref{sec:min}. This is the viscous relaxation scheme \citep{bajer2013} where we again solve the induction equation \eqref{eqn:induc}, but determine $\bm{u}$ from the force balance
\begin{equation}
    \mu_s\Delta\bm{u} + \left(\frac13\mu_s + \mu_b\right){\bm\nabla}({\bm\nabla}{\bm\cdot}\bm{u}) + \bm{J}\times\bm{B} = \bm{0}.
\end{equation}
In our one-dimensional case this reduces to
\begin{equation}
    \mu\frac{\upartial{}^2u_x}{\upartial{x}^2} - \frac{\upartial{}}{\upartial{x}}\left(\frac{B_y^2}{2\mu_0}\right) = 0,
    \label{eqn:visc1d}
\end{equation}
where $\mu = 4\mu_s/3 + \mu_b$. For our periodic system, this scheme again has the property that the magnetic energy decreases monotonically, with
\begin{equation}
    \frac{\mathrm{d}}{\mathrm{d}t}\int_V\frac{B^2}{2\mu_0}\,\mathrm{d}V = -\mu\int_V\left|\frac{\upartial{u_x}}{\upartial{x}}\right|^2\,\mathrm{d}V.
    \label{eqn:Wnu}
\end{equation}
In the periodic system, equation \eqref{eqn:visc1d} specifies $u_x$ only up to an additive constant. In view of the symmetry of our test problem, we set $u_x(-1/2, t) = 0$, so solving \eqref{eqn:visc1d} for $u_x$ gives
\begin{equation}
    u_x(x,t) = \frac{1}{\mu}\int_{-1/2}^x\frac{B_y^2}{2\mu_0}\,\mathrm{d}x - \frac{x+1/2}{2\mu}\int_{-1}^1\frac{B_y^2}{2\mu_0}\,\mathrm{d}x.
\end{equation}
Substituting this into the induction equation \eqref{eqn:ind} gives an integro-differential evolution equation for $B_y$. Figure~\ref{fig:visc} shows a numerical solution on the same uniform mesh used for the MF simulations, but using a simple upwind scheme (as the equation is now hyperbolic). Unlike MF, we see that the viscous relaxation method evolves rapidly toward the minimum energy state. In future, it would be fruitful to explore as an alternative to MF in the context of the solar coronal magnetic field, where null points are prevalent. 

\section*{Acknowledgements}

This work was supported by UKRI/STFC research grant ST/S000321/1. The author thanks the two anonymous referees for improving the paper.

\bibliographystyle{gGAF}

\appendices
\section{Numerical method for MF} \label{sec:app}

The MF equation \eqref{eqn:mf1d} has the form of a nonlinear diffusion equation, and is either parabolic or degenerate parabolic, depending on the form of $\nu$. As such it is best treated with an implicit method, although solar physics applications have often used an explicit method with the resulting timestep restriction. In this paper we adopt an implicit Crank-Nicolson scheme, treating the nonlinearity with Picard iteration. For shorthand, let $B=B_y$ and $\beta(B^2) = B^2/[\mu_0\nu(B^2)]$, so that the equation is
\begin{equation}
    \frac{\upartial{B}}{\upartial{t}} = \frac{\upartial{}}{\upartial{x}}\left(\beta(B^2)\frac{\upartial{B}}{\upartial{x}}\right).
\end{equation}
We take a uniform mesh $x_j = j\Delta x -1$ for $j=0,\cdots, n_x-1$, with spacing $\Delta x = 2/(n_x-1)$. The magnetic field values $B_{j+\frac12}^n$ are located at cell centres $\frac12(x_j + x_{j+1})$, where $n$ denotes the time level $t_n = n\Delta t$. A single timestep, to compute $B_{j+\frac12}^{n+1}$ from $B_{j+\frac12}^{n}$, involves iteratively solving the Crank-Nicolson formula over $k$, where
\begin{align}
    &-\alpha_{j}^{n,k}B_{j-\frac12}^{n, k+1} + \left(1 + \alpha_{j}^{n,k} + \alpha_{j+1}^{n,k}\right)B_{j+\frac12}^{n,k+1} - \alpha_{j+1}^{n,k}B_{j+\frac32}^{n,k+1} \nonumber \\
    & \hskip 40mm = \alpha_{j}^{n,0}B_{j-\frac12}^{n} + \left(1 - \alpha_{j}^{n,0} - \alpha_{j+1}^{n,0}\right)B_{j+\frac12}^{n} + \alpha_{j+1}^{n,0}B_{j+\frac32}^{n},\hskip 10mm
    \label{eqn:crank}
\end{align}
starting from $B_{j+\frac12}^{n,0} =B_{j+\frac12}^{n}$. The coefficients comprise the scaled ``diffusivity'' $\beta$ evaluated at the mesh points,
\begin{subequations}
\be
\alpha_j^{n,k} = \frac{\Delta t}{2\Delta x^2}\beta\left(\bigl[B^2\bigr]_j^{n,k}\right),
\ee
in which
\be
\bigl[B^2\bigr]_j^{n,k} = \frac14\biggl[\Bigl(B_{j-\frac12}^n\Bigr)^{\!2} + \Bigl(B_{j+\frac12}^n\Bigr)^{\!2} + \Bigl(B_{j-\frac12}^{n,k}\Bigr)^{\!2} + \Bigl(B_{j+\frac12}^{n,k}\Bigr)^{\!2}\biggr].
\ee
\end{subequations}
Equation \eqref{eqn:crank} is iteratively solved until $\left\|B_{j+\frac12}^{n,k+1} - B_{j+\frac12}^{n,k}\right\|_\infty < 10^{-14}$, then we set \hfill\break
$B_{j+\frac12}^{n+1} = B_{j+\frac12}^{n,k+1}$.

We have verified the convergence against exact solutions for (i) linear diffusion when $\nu = B^2$, and (ii) ambipolar diffusion when $\nu=1$. For (ii) we do not have the exact solution for our test problem, so we tested instead with the well-known Barenblatt solution \citep{vazquez2006} in the form
\begin{equation}
    B_y(x,t) = 
    \sqrt{\max\left\{0, \frac{1}{10}\tilde{t}^{-1/2} - \frac{x^2}{12}\tilde{t}^{-1} \right\}}, \hskip 10mm \tilde{t} = \dfrac13\left(t + \dfrac{1}{10}\right).
\end{equation}
For case (i), and in smooth regions of case (ii), the expected second-order convergence is recovered, where we took $\Delta t = 10\Delta x^2$. The position of the moving ``front'' at $x=(\sqrt{6/5})\tilde{t}^{1/4}$, where the slope of $B_y$ in the exact solution becomes infinite and discontinuous, still converges, but only to first order.

\end{document}